\documentclass[preprint,amsmath,amssymb,aps,superscriptaddress]{revtex4-1}
\usepackage{graphicx}
\usepackage{enumitem}
\usepackage{xcolor}
\usepackage{float}
\usepackage{enumitem}
\usepackage{bm}

\def\h{hydrogen} \def\vdis{{\bar v}_{\rm dis}}
\def\aFe{$\alpha$-Fe} 
\def\dP{h} 
\def\P{x} 
\def\Ekp{E_{\rm kp}} 
\def\occ{\chi} 
\def\hocc{\occ_i} 
\def\hoccn{C_0} 
\def\hocctc{\occ_t} 
\def\hocce{\occ^{\rm e}_i} 
\def\CH{C_{\rm H}} 
\def\i{\textit{i\/}} \def\ii{\textit{ii\/}} \def\iii{\textit{iii\/}} \def\iv{\textit{iv\/}} \def\v{\textit{v\/}}

\begin{document}
\title{The influence of hydrogen on plasticity in pure iron---theory and experiment}
\author{Peng Gong}
\affiliation{Department of Materials Science and Engineering, University of Sheffield, Mappin Street, Sheffield, S1 3JD, UK}
\author{Ivaylo H.~Katzarov}
\affiliation{Department of Physics, King's College London, Strand, London, WC2R 2LS, UK}
\affiliation{Bulgarian Academy of Science, Institute of Metal Science, 67, Shipchenski prohod Str., 1574 Sofia, Bulgaria}
\author{Anthony T.~Paxton}
\affiliation{Department of Physics, King's College London, Strand, London, WC2R 2LS, UK}
\author{W.~Mark Rainforth}
\affiliation{Department of Materials Science and Engineering, University of Sheffield, Mappin Street, Sheffield, S1 3JD, UK}

\date{\today}

\begin{abstract}
Tensile stress relaxation is combined with transmission electron microscopy to reveal dramatic changes in dislocation structure and sub structure in pure \aFe\ as a result of the effects of dissolved \h. We find that \h\ charged specimens after plastic deformation display a very characteristic pattern of trailing dipoles and prismatic loops which are absent in uncharged pure metal. We explain these observations by use of a new self consistent kinetic Monte Carlo model, which in fact was initially used to \textit{predict} the now observed microstructure. The results of this combined theory and experimental study is to shed light on the fundamental mechanism of \h\ enhanced localised plasticity.
\end{abstract}

\maketitle

\section{Introduction}

The subject of \h\ influence on the mechanical behaviour of steel is hugely controversial. On the other hand the dramatic effects of \h\ on the mechanical integrity of engineering structures is well documented~\cite{Telegraph2015} and if society is to enter a future hydrogen economy the problem must be tackled head on. The violent reduction in fracture toughness of steel as a consequence of dissolved \h\ at the level of some atomic parts per million (appm), at the broadest level of current understanding is either the result of a loss of cohesive strength (the HEDE hypothesis) or the consequence of enhanced localised plasticity (the HELP hypothesis). Other theories such as the role of accumulated vacancy damage or the emission of dislocations from crack surfaces have also been proposed~\cite{Simpson2017}. One of the striking features of the problem has been a lack of detailed confirmation of observation with theory and modelling; and vice versa. A particular difficulty arises from the putative elastic shielding of dislocation strain fields due to hydrogen. This is well documented both in elegant electron microscopy observations~\cite{Ferriera1988,Sofronis2002} and sophisticated theoretical treatments in linear elasticity~\cite{Birnbaum1994,Sofronis1995}. The proposal that HELP is a consequence of elastic shielding by Cottrell atmospheres is untenable in steel because the solubility of \h\ in body centred cubic \aFe\ is about six orders of magnitude too small for the effect to be measurable~\cite{Ferriera1988,Song2014}. Conversely it has been proposed that the \h\ trapped locally in the cores of dislocations is responsible for the enhanced plasticity~\cite{Yu2019}. Furthermore it is not obvious that \h\ will increase dislocation mobility under all circumstances. In fact \h\ may increase or decrease dislocation velocity depending on the conditions of \h\ concentration, temperature and applied stress in pure \aFe~\cite{iky2}. 

Here, we present for the first time a self consistent kinetic Monte Carlo model that is able to predict average dislocation velocity and to simulate microstructural development that arises from \h\ self pinning effects. We confirm predictions of the model by transmission electron microscopy (TEM) observations. In both cases we use pure \aFe, and we match as closely as possible the experimental and modelling conditions. Furthermore we make contact between our calculations and recently published measurements of activation volume~\cite{Wang2013} and we find a striking accord between experiment and theory. Finally, we conclude with speculations about the role of \h\ in the generation of dislocation cellular structure which make contact with modern theories of work hardening~\cite{Brown2012}.

\section{Theoretical}
\label{section_theory}
\subsection{Introduction and background}

At the heart of the simulation of \h\ effects on plasticity is the model that is used to describe the connection between background, or nominal, \h\ concentration, $\CH$, here defined in units of atomic parts per million, appm; and either the flow stress or the average dislocation velocity, $\vdis$. At the simplest level as used in typical discrete dislocation dynamics simulations or crystal plasticity finite element models, simple ad hoc assumptions are used~\cite{Barera2013,Castelluccio2018}. However, $\vdis$ is a complex function of $\CH$, and depending on applied stress and temperature $\vdis$ can be both \textit{enhanced} and \textit{reduced} depending on the background \h\ concentration~\cite{KPP2017}.  In earlier work~\cite{KPP2017,Footnote1}, two of us developed an off-lattice kinetic Monte-Carlo method to calculate the velocity of screw dislocations in \aFe\ based upon first principles calculations of kink-pair formation energies~\cite{iky2}. In this model a number of quite serious approximations are made, namely, (\i) the kink pair formation energy is affected only by \h\ ahead of the dislocation in the glide plane; (\ii) kink velocity is only affected by \h\ behind the dislocation in the glide plane; (\iii) \h\ is assumed to remain fixed in place during kink-pair formation and migration; (\iv) the time for segments of dislocation to move between Peierls valleys is assumed greater than the \h\ jump time within the dislocation core. In spite of its simplicity that model was able to predict dislocation velocity as a function of temperature, stress, $\tau$, and nominal \h\ concentration and it was shown that such a function is non monotonic and that the effect of \h\ can be to increase or decrease dislocation velocity, depending on conditions; and that the change in velocity compared to pure \aFe\ at 300K and $\tau=100$~MPa increases by more than a factor of 10 up to 5~appm and then decreases to less than the velocity in pure \aFe\ at 20~appm. In addition, the simulations~\cite{KPP2017} \textit{predicted} that under most conditions of \h-loaded \aFe\ a moving screw dislocation will leave a trail of debris made up of rows of prismatic loops. The central result of the present paper is that \textit{we have found these loops} in electron microscope images of deformed, \h-charged \aFe. On the other hand the model was not able to reproduce activation volume measurements~\cite{Wang2013} which indicate a minimum in the dislocation velocity as a function of hydrogen concentration at about 10~appm at 300K. We present here a new model which we call ``self consistent kinetic Monte-Carlo'' (SCkMC) which permits a dynamic non equilibrium distribution of \h\ about the moving dislocation core. Specific new features of the model are, (\i) simultaneous kink nucleation, migration and \h\ jumping; (\ii) kink pair formation energy affected by all \h\ within the core; (\iii) a non equilibrium distribution of \h\ which depends on temperature and average dislocation velocity; (\iv) kink pair formation energy depends on average dislocation velocity; (\v) mobile \h\ during glide---although the \textit{total} \h\ occupancy within the core is assumed fixed.

\subsection{Line tension model}
\label{section_LT}

The SCkMC is predicated on a parameterised line tension model~\cite{iky1,iky2}. We imagine a long dislocation lying in its Peierls valley, a segment of which has migrated towards or into the next Peierls valley so as to make an incipient or complete kink pair. The dislocation is divided into bins of width $b$, the Burgers vector, along its length and a variable ${\P}_j$ is assigned to describe the deviation of the segment lying in the $j^{\rm th}$ bin from the dislocation's original position in the Peierls valley---the elastic center of the dislocation. There is a periodic Peierls energy landscape described by an energy function, $E_p\left({\P}_{j}\right)$. The energy per unit length of dislocation is then prescribed in the following line tension expression~\cite{iky2},
\begin{align}
E &= \sum_j E_j \nonumber\\
  &= \frac{1}{2} K\sum_{j}\left({\P}_{j}-{\P}_{j+1}\right)^2
  + \sum_{j} E_p\left({\P}_{j}\right) + \sum_{j}\epsilon_{1pq}\tau_{pr}b_{r}\xi_{p}\P_{j}
  -\sum_{jk}E_{\rm H}\left(\left\vert{\P}_{j}-{\P}^{\rm H}_{k}\right\vert\right) 
\label{eq_LT}
\end{align}
The first term describes the energy penalty for two bins which have different amounts of deviation from the original Peierls valley towards the next and $K$ is the associated ``spring constant''. The second term is the energy of the segment $j$ depending on its height in the Peierls landscape. The third term, with an implicit sum from~1 to~3 over $\{pqr\}$, is the 1-component (perpendicular to $[111]$) of the  Peach--Kohler force arising from a local stress $\tau_{pq}$ times the displacement of the $j^{\rm th}$ segment of dislocation having a line sense $\bm{\xi}$. This term ``tilts'' the corrugated energy landscape so that the Peierls valley ahead of the dislocation is lower in energy that the one behind, and provides the driving force for glide. The final term expresses the energy associated with a \h\ atom that is trapped at a position at a distance $\vert{\P}_{j}-{\P}^{\rm H}_{k}\vert$ from the core, in which ${\P}^{\rm H}_{k}$ is the position of the $k^{\rm th}$ \h\ atom relative to the elastic centre.

\subsection{Dynamics of the long straight dislocation}

We first examine the motion of a long straight dislocation, its line moving as a whole. And in the next section we address the actual situation of glide by the Peierls mechanism of kink pair creation and kink migration~\cite{Caillard2010}.
Density functional theory (DFT) calculations have identified two core structures of the $\frac{1}{2}[111]$ screw dislocation, the so called ``easy core'' (EC), which is the stable, low energy configuration, and the ``hard core'' (HC) which is metastable~\cite{Clouet2009,iky1}. The HC is very close in configuration to the ``saddle point'' (SP) core~\cite{Clouet2009,iky1}. DFT calculations furthermore show that \h\ binds strongly to the EC with three equivalent sites having binding energies of $E_i=256$~meV in the so called $E_1$/$E_2$ basin, three in the $E_3$/$E_4$ basin having $E_i=201$~meV and six in the $E_7$/$E_8$ basin with $E_i=77$~meV~\cite{iky2}. The strongest binding sites for the HC are one in the $H_0$/$H_1$ basin located at the centre of the core with $E_i=390$~meV, and six binding sites denoted $H_2$ having $E_i=189$~meV~\cite{iky2}. As a dislocation moves from EC to HC to EC the $E_1$/$E_2$ traps sites ahead of the dislocation line transform into $H_0$/$H_1$ sites and finally the \h\ occupies $E_1$/$E_2$ traps sites behind the dislocation line.

When the dislocation is lying in its equilibrium Peierls valley the probability of occupancy of a trap site, $i$, is determined by the McClean isotherm,
\begin{equation}
    \hocc = \frac{\frac{1}{6}\hoccn\, e^{E_i/kT} }
             {1 + \frac{1}{6}\hoccn\, e^{E_i/kT} }
\label{eq_McLean}
\end{equation}
in which $\hoccn=10^{-6}\CH$ is the nominal number of \h\ atoms per Fe atom, and the factor $1/6$ accounts for there being six tetrahedral sites per bulk Fe atom. Here, $k$ is the Boltzmann constant and $T$ is the absolute temperature. If we take a sum over all the trap sites in the dislocation core, we will define 
\begin{equation}
\hocctc = \sum_i \hocc = \hbox{constant}
\label{eq_Htotal}
\end{equation}
as the total \h\ occupancy of the core sites; and we will assume throughout that \textit{this is constant}, that is, \h\ will redistribute dynamically between trap sites during glide but overall the dislocation will not absorb or reject \h; we also only allow \h\ to redistribute among traps within a plane perpendicular to the dislocation line, in view of the very slow \h\ pipe diffusivity~\cite{Bombac2017}. In the case of slow glide, and the maintenance of equilibrium, then as a long straight dislocation moves between two Peierls valleys, we may define the occupation probability of trap site $i$ as~\cite{HirthLothe},
\begin{equation}
\hocce(x) = \frac{\hocctc e^{-E_i(x)/kT}}{\sum_j e^{-E_i(x)/kT}} 
\label{eq_occ}
\end{equation}
Here $0<x<\dP$, if $\dP=a\sqrt{2/3}=2.34$~\AA\ is the period of the Peierls potential on the $(\bar 110)$ plane, so that $x$ describes the position of the dislocation line with respect an origin at the EC elastic centre. As the dislocation glides \h\ will redistribute between trap sites, which themselves distort and therefore whose trap depth, $E_i(x)$, varies with $x$. We parameterise $E_i(x)$ by fitting and interpolation of DFT data~\cite{iky2}. Once that is done, then in association with the line tension model~(\ref{eq_LT}) we have a complete description of the energetics of the dislocation as a function of $x$ and the total occupancy, $\hocctc$, \textit{for the moment only} in two limiting cases: (a)~equilibrium, slow glide in which traps are occupied according to~(\ref{eq_occ}), and (b)~fast glide, in which all \h\ atoms are fixed in the traps they occupy in the EC initial state before glide. 

\begin{figure}
    \caption{Peierls potential: the potential energy in units of eV per Burgers vector of a long straight $\frac{1}{2}[ 111\rangle$ screw dislocation as a function of distance between one Peierls valley and the next. (a) Limiting case of slow motion: the \h\ remains in equilibrium and moves reversibly between $E_1$/$E_2$ basins. At the saddle point the \h\ is trapped at the $H_0$/$H_1$ basin near the saddle point. Note, how as \h\ concentration is increased above 30~appm the saddle point core structure becomes more stable than the easy core. (b) Limiting case of high dislocation velocity: the \h\ remains behind in a trap site of high energy compared to the $E_1$/$E_2$ basin hence the line tension is greater after glide by one repeat distance than before. The curves are labelled with the nominal background \h\ concentration, $\CH$. Temperature is 300K.}
    \centering
    \includegraphics[scale=0.75,viewport=13 115 392 347,clip]{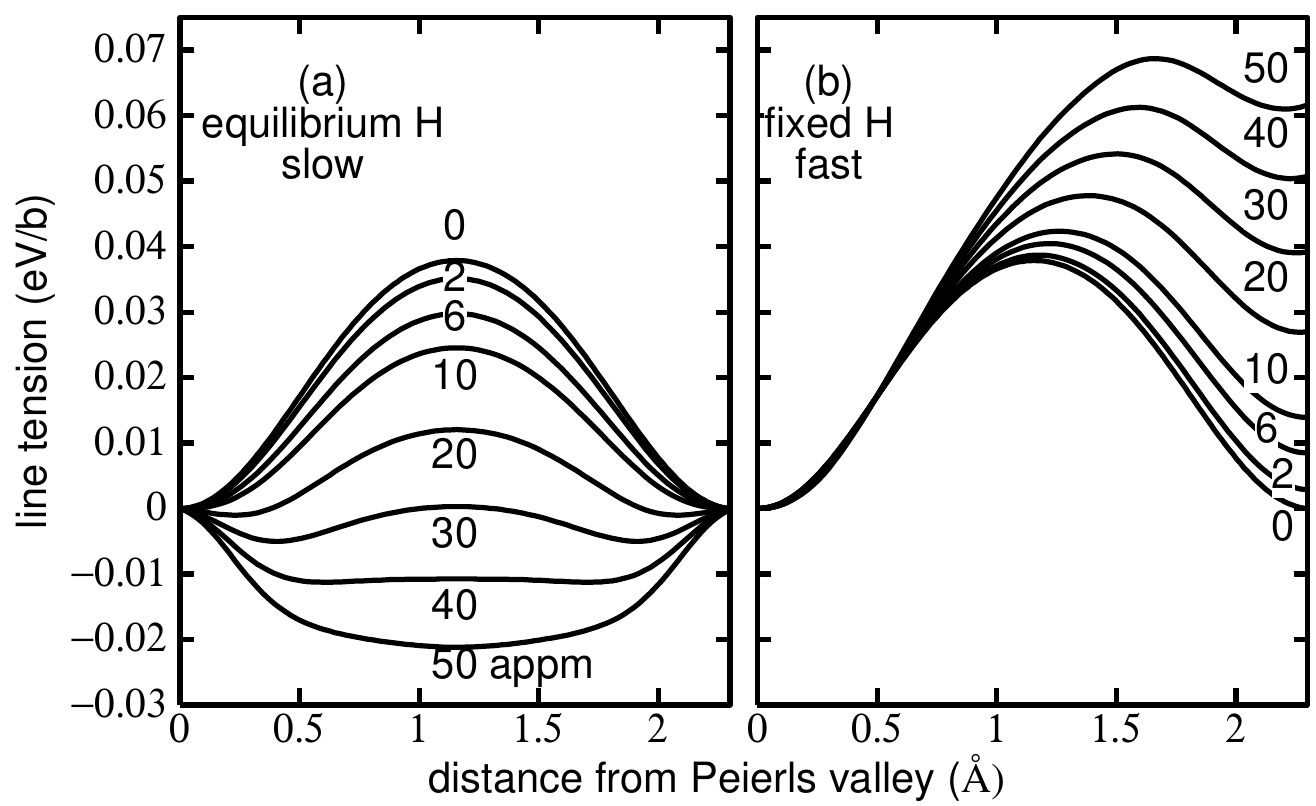}
    \label{fig:LineTension}
\end{figure}

\begin{description}[topsep=0pt, partopsep=0pt, itemsep=0pt, listparindent=12pt, itemindent=12pt, parsep=0pt, leftmargin=6pt, partopsep=12pt, labelwidth=12pt]
\item [(a)] Fig.~\ref{fig:LineTension}(a) shows potential energy profiles in the equilibrium limit of a slowly moving dislocation. At $\CH=0$ the profile is typical of a calculated Peierls barrier~\cite{Mrovec2011}. The Peierls barrier shown predicted by our model is consistent with the measured estimate of 37~meV/b~\cite{Caillard2010}. The barrier becomes smaller as $\CH$ is increased because \h\ is stablising the saddle point core as the $E1/E2$ traps distort into $H0/H1$ traps. In fact the effect is strong enough so that when $\CH$ exceeds 30~appm the saddle point core is lower in energy than the easy core and their roles are reversed; this is because the total energy gained by \h\ in deeper traps overwhelms the penalty in core energy. In this way the Peierls barrier is reduced to close to zero and then increases again. However above about 30~appm \h\ the saddle point is at the EC, and the minimum is in the HC configuration. 
\item [(b)] In the limit of \textit{rapid glide} the \h\ atoms are kept fixed during the movement of a dislocation between Peierls valleys then as the dislocation moves, \h\ that was trapped in deep traps may not jump into the newly created traps, but instead remains behind in sites of higher potential energy; hence the Peierls barrier increases continually with $\CH$ and the initial and final positions of the dislocation line have not the same energy: the profile is asymmetric as shown in Fig.~\ref{fig:LineTension}(b).
\end{description}

A highly relevant conclusion is that trapped \h\ serves to \textit{stabilise the hard core with respect to the easy core}, so that \h\ is able to trigger a core transformation which strongly modifies the Peierls barrier.

\begin{figure}
    \caption{Potential energy of a long straight $\frac{1}{2}[111]$ screw dislocation as in Fig.~\ref{fig:LineTension}. Panels (a), (b) and (c) show solutions using the continuity equation~(\ref{eq_cont}) at nominal \h\ concentrations of 10, 30 and 50~appm respectively. In cases where the saddle point is of lower energy than the EC end points, the Peierls barrier is inferred by taking the end points as saddle point energies and the stable core to be the HC. Curves for dislocation velocities, $v$, between $5\times 10^4$ and $3\times 10^{10}$~nm~s$^{-1}$ are shown. Temperature is 300K.}
    \centering
    \includegraphics[scale=0.75,viewport=12 115 558 348,clip]{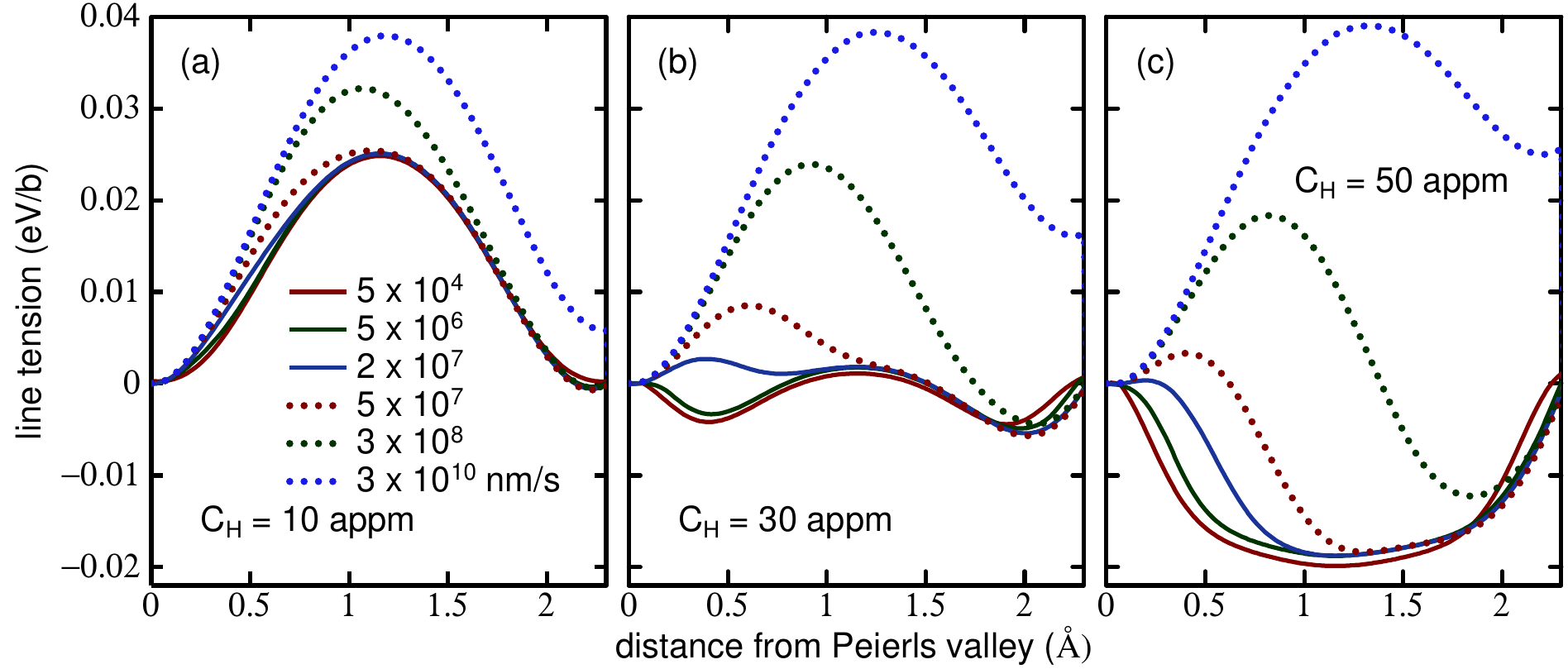}
    \label{fig:LineTensionHydrogen}
\end{figure}

The results in Fig.~\ref{fig:LineTension} suggest to us that the actual profile will be somewhere in between the two limits, the departure from equilibrium being controlled by the uniform dislocation velocity, $v$. Therefore we seek a theory that will predict the profile as a function of $v$. Because of the finite speed of the dislocation, we expect that the probability of occupancy of trap $i$, $\hocc(x)$, differs from its equilibrium value~(\ref{eq_occ}), $\hocce(x)$. For any of the 10 strongest binding sites we find the following continuity equation,
\begin{equation}
\frac{\partial \hocc(x,v)}{\partial t} = v\,\frac{\partial \hocc(x,v)}{\partial x}=
\left(\hocce(x) - \hocc(x,v)\right)\,\nu\,{\rm e}^{-E_i(r)/kT}
\label{eq_cont}
\end{equation}
where $\nu$ is an ``attempt frequency'' for \h\ to escape from the $i^{\rm th}$ trap~\cite{Paxton2016}. By solving~(\ref{eq_cont}), subject to the condition~(\ref{eq_Htotal}) that the total \h\ occupancy remains constant, we may determine the potential energy of the dislocation as a function of position between two Peierls valleys at velocity, $v$. We show these data in Fig.~\ref{fig:LineTensionHydrogen}.

We observe that at the critical $\CH$ of 30~appm where the Peierls barrier for low velocity is close to zero, the actual barrier is strongly dependent on the velocity and only vanishes in the slow, equilibrium limit.

\subsection{Dynamics by kink pair creation and migration}

The $\frac{1}{2}[111]$ screw dislocation in bcc transition metals is characterised by its non planar, non degenerate core structure~\cite{Groger1} which means that even at the lowest temperatures, its glide is via a \textit{Peierls mechanism}, namely the process of kink pair creation followed by kink migration~\cite{HirthLothe}. Kink pair generation is thermally activated. We therefore turn now to the actual problem of predicting $\vdis$ within the Peierls mechanism~\cite{Caillard2010}. 

\subsubsection{Kink pair creation}

The screw dislocation does not lie quiescent in its Peierls valley; fluctuations produce random events in which a small section deviates towards a neighbouring Peierls valley. Mostly this produces an ``incipient'' kink pair which annihilates due to elastic attraction of the kinks. A stable kink pair is one that has sufficient distance between the kinks, which we take to be about $30b$~\cite{iky1,KPP2017}, that elastic attraction is small enough to allow the kink pair to survive and its halves to separate under the local stresses they encounter. The formation of a stable kink pair is a result of numerous acts of kink-pair nucleation, annihilation, and increasing distance between kinks under the action of the applied shear stress. We do not consider all these processes explicitly in our simulations. The rare event of formation of a stable kink pair, which separates under the resolved shear stress is treated using the kinetic Monte-Carlo procedure described elsewhere~\cite{Deo2005,BulatovCai,KPP2017}.

\begin{figure}
    \caption{Kink pair formation enthalpy, $\Ekp$, as a function of resolved shear stress, $\tau$ and average dislocation velocity, $\vdis$; calculated by iterative solution of equations~(\ref{eq_vconst}) and~(\ref{eq_tr}).}
    \centering
    \includegraphics[scale=0.75,viewport=15 10 445 340,clip]{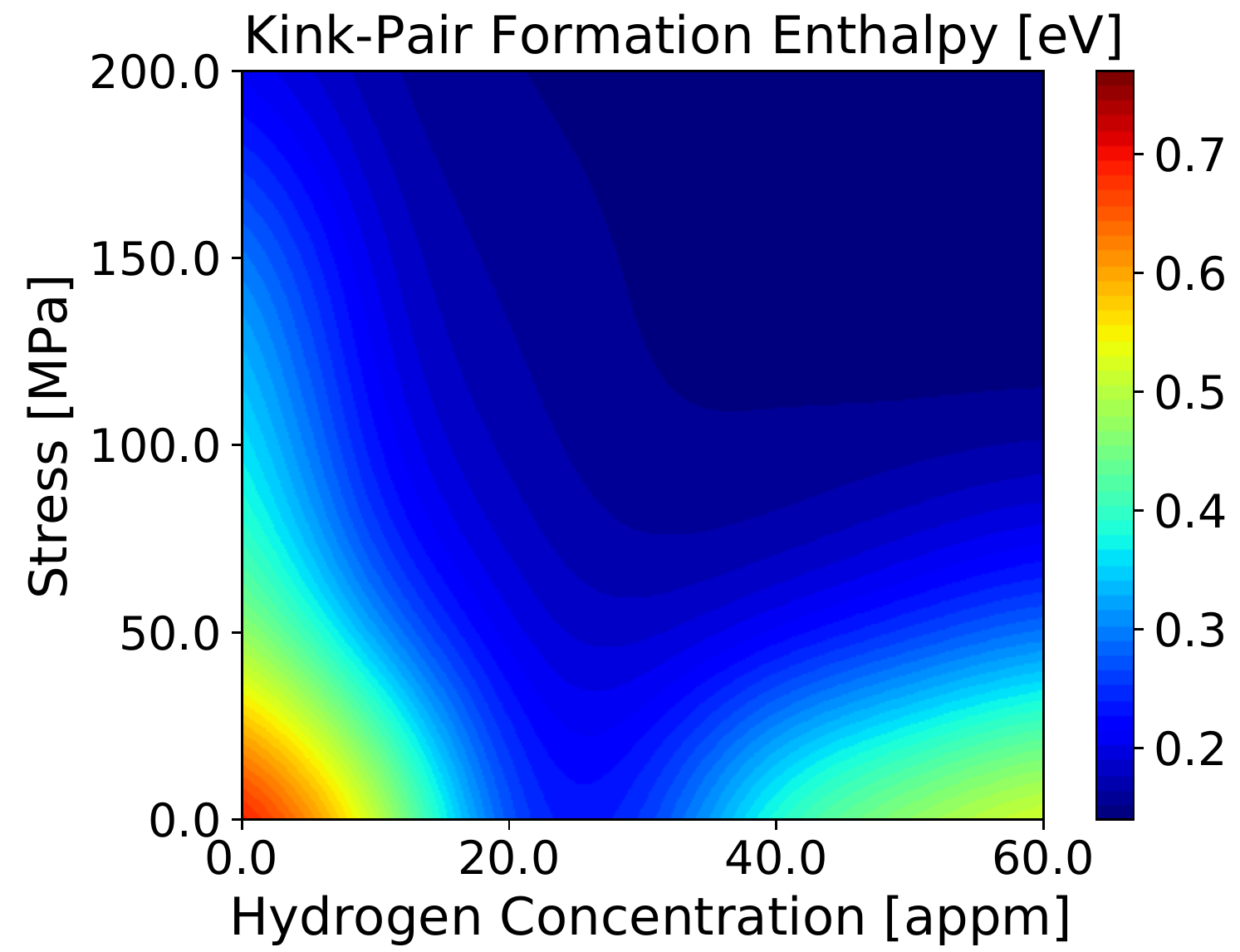}
    \label{fig:EkpVelStress}
\end{figure}

The reason for requiring a \textit{self consistent} theory is that trapped \h\ will strongly modify the kink pair formation enthalpy, $\Ekp$, and that the location of \h\ in traps will depend on how fast the dislocation is moving. Hence $\Ekp$ is a function of $\vdis$ since it depends on the rate at which \h\ is distributed among trap sites as the dislocation glides. For a given resolved shear stress, $\tau$, and an assumed average velocity, $\vdis$, using the line tension model and data such as in Fig~\ref{fig:LineTensionHydrogen} the energy, $E_j(\CH,x,\vdis)$, of a dislocation segment, Eq.~(\ref{eq_LT}), of length $b$ and at a distance $x$ from the EC elastic centre in the initial Peierls valley, can be calculated. Then using linear, non-singular elastic theory~\cite{HirthLothe,BulatovCai} and the ``nudged elastic band'' (NEB) method~\cite{Henkelman2000}, we may calculate the kink pair formation enthalpy, $\Ekp(\CH,\tau,\vdis)$. \textit{However} $\Ekp$ is a function of $\vdis$ while $\vdis$ is a function of $\Ekp$: $\Ekp=\Ekp(\vdis)$ and $\vdis=\vdis(\Ekp)$. To make progress and to find a self consistent solution, we assume that the average speed is constant, and
\begin{equation}
\vdis(\Ekp)=\frac{\dP}{t_r}
\label{eq_vconst}
\end{equation}
allowing us to define an \textit{average relaxation time} for kink pair formation,
\begin{equation}
    t_r=\nu_{\rm kp}^{-1}{\rm e}^{\Ekp(\CH,\tau,v)/kT} 
\label{eq_tr}
\end{equation}
where $\nu_{\rm kp}$ is an attempt frequency for which we use the Debye frequency of \aFe. 
In order to solve~(\ref{eq_vconst}) and~(\ref{eq_tr}), and to determine $\Ekp$ at given $\CH$ and $\tau$, we proceed with the following iterative process.
\begin{description}[topsep=0pt, partopsep=12pt, itemsep=0pt, listparindent=12pt, itemindent=12pt, parsep=0pt, leftmargin=6pt, partopsep=12pt, labelwidth=12pt, labelsep=0pt]
\item[1.] Assume an initial $\Ekp$.
\item[2.] Calculate the corresponding $\vdis$ using~(\ref{eq_vconst}) and~(\ref{eq_tr}).
\item[3.] Determine the distribution of \h\ from the continuity equation~(\ref{eq_cont}), subject to~(\ref{eq_Htotal}); and calculate the segment energy, $E(\CH,x,\vdis)$ from the line tension model.
\item[4.] Calculate $\Ekp$ using the NEB and go to step 2.
\end{description}
This process is iterated until $\Ekp$ calculated in step~4 is no longer changing to within some tolerance. Figure~\ref{fig:EkpVelStress} shows the results of the iterative procedure. 

We may interpret Fig.~\ref{fig:EkpVelStress} in the following way. At high stress, $\Ekp$ is uniformly small because the applied stress acts to drive a dislocation into the next Peierls valley and this dominates the process of glide. At low stress we observe a large $\Ekp$ at low $\CH$, the largest being that of pure \aFe\ and zero stress. As $\CH$ increases, $\Ekp$ decreases, consistent with the calculations shown in Fig.~\ref{fig:LineTension}. $\Ekp$ reaches a minimum at $\CH\approx 30$~appm in Fig.~\ref{fig:EkpVelStress} as predicted in Fig.~\ref{fig:LineTension} and this minimum in $\Ekp$ is a consequence of the \h-induced core transformation from EC to HC. As $\CH$ increases further $\Ekp$ rises as a consequence of the increasing Peierls barrier---but now the barrier is at the easy core configuration and the Peierls valley corresponds to the HC.

\subsubsection{Kink migration}

Glide is a two step process. After the formation of a stable double kink the two kinks will separate in opposite directions. In pure metal, the kink migration or \textit{secondary Peierls barrier} is low and is not thermally activated. However \h\ and other interstitials change that. If a \h\ atom is trapped in the $E1/E2$ basin just behind the dislocation line and a kink sweeps past, then that \h\ ends up in a higher enthalpy trap site~\cite{iky2}, which implies that thermal activation is then required for the kink to proceed. We do not need to rehearse the kMC procedure here since we use the identical scheme as described earlier~\cite{KPP2017}. However we should underline the \textit{physics} here since it is essential in appreciating the present findings. In the case of pure \aFe, a screw dislocation of typical length of about $1000b$ will glide as a unit as in face centred cubic metals (albeit by thermal activation of kink pairs) since the kink migration speed is so fast that a kink pair has separated to the ends of the dislocation before the next kink pair is activated~\cite{KPP2017}. Hence kink collision does not occur. The situation is very different if the kinks suffer solute drag due to \h\ and other interstitials. A key fact is that kink pairs are created on any one of the three $\lbrace\bar 110\rbrace$ glide planes in the zone of the $[111]$ Burgers vector. \textit{If two kinks on different glide planes collide} the resulting defect is an edge jog which is sessile. Our findings earlier~\cite{KPP2017}, which we confirm here, are that such jogs amount to self pinning points which drag out edge dipoles and these dipoles will pinch out to create a train of prismatic loop debris, \textit{entirely as a consequence of dissolved \h.}

\subsection{Results of the self consistent kinetic Monte-Carlo simulations}

\subsubsection{Average dislocation velocity}

\begin{figure}
    \caption{Average dislocation velocity calculated within the self consistent kinetic Monte-Carlo model using $\Ekp$ from Fig.~\ref{fig:EkpVelStress}.}
    \centering
    \includegraphics[scale=0.75,viewport=9 18 442 334,clip]{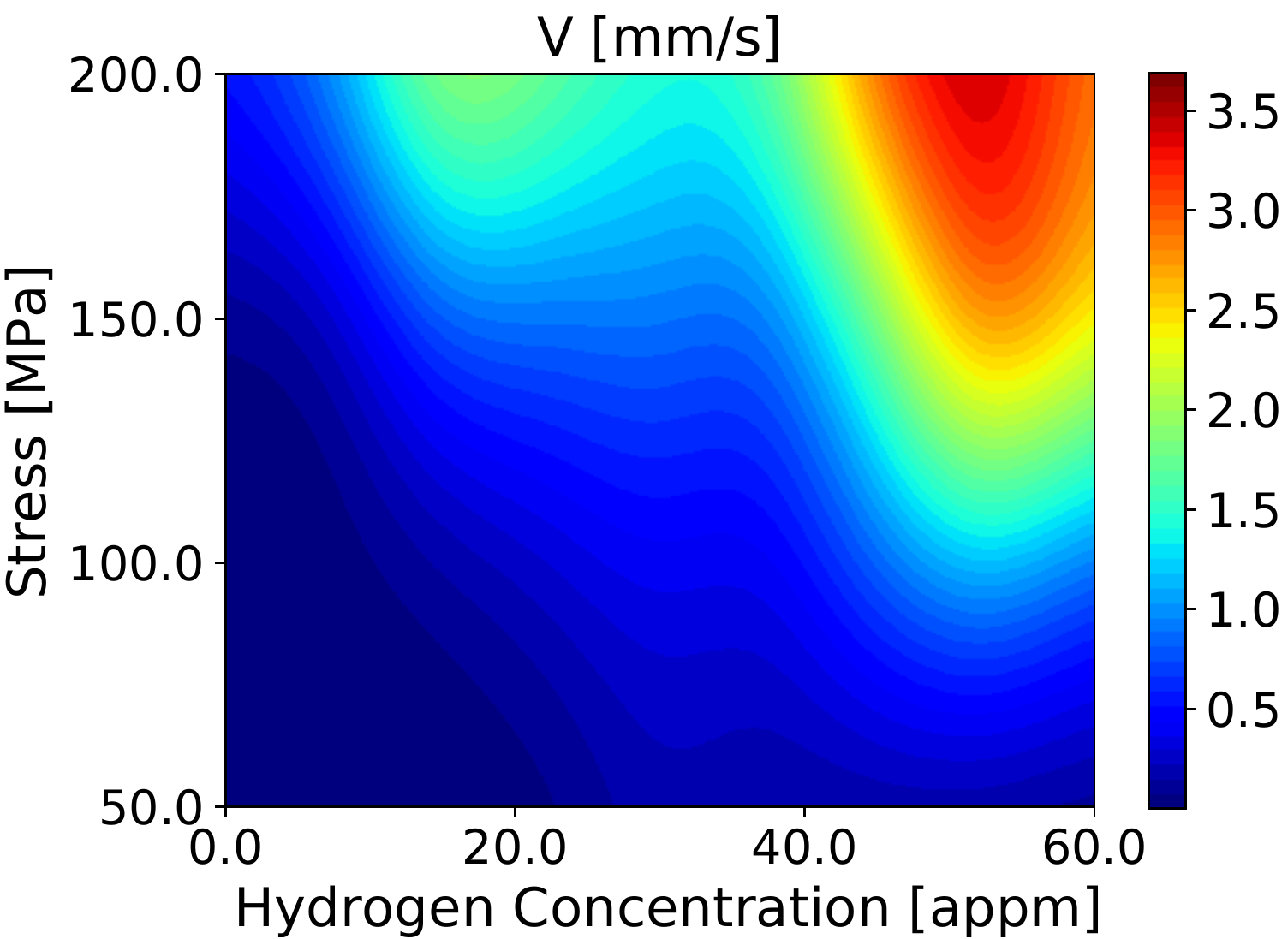}
    \label{fig:VelHStress}
\end{figure}

Conditions of the present self consistent kMC simulations are identical to those of the earlier non self consistent modelling~\cite{KPP2017}. Temperature is 300K. Rather than using a kink pair formation energy that depends only on stress, $\CH$ and temperature; we now employ $\Ekp$ as a function of $\vdis$ also as taken from Fig~\ref{fig:EkpVelStress}. Average dislocation velocity as a function of stress and $\CH$ is shown in Fig.~\ref{fig:VelHStress}.

For $\tau < 50$~MPa (not shown in Fig.~\ref{fig:VelHStress}), $\vdis$ increases with $\CH$, reaching a maximum at $\CH\approx 25$~appm; thereafter $\vdis$ decreases as a result of the increase in $\Ekp$ (Fig.~\ref{fig:EkpVelStress}). At $\tau > 100$~MPa, $\vdis$ does not go through a minimum, but increases steadily with $\CH$ until $\CH\approx 40$~appm at which a rather dramatic increase is found, followed by a decline at higher \h\ concentrations. The greatest average dislocation velocity, for all stresses, occurs at a nominal \h\ concentration of about 50~appm. This complex behaviour can be traced in part to the concentration dependence of the kink pair formation enthalpy and the \h-induced core transition from easy core to hard core. If $\Ekp$ is small or vanishing then kink pair formation is easy on all three glide planes in the zone of the Burgers vector, and this leads to increased likelihood of kink pair collisions on different glide planes. Once an immobile jog is created further kinks pile into it, resulting in the formation of superjogs and trailing dislocation dipoles (see Fig.~\ref{fig:jogs}. The two arms of the dipole may intersect and recombine by kink pair recombination. Thereby the dipole is unzipped and a straight dislocation in screw orientation is restored. This involved set of operations serves greatly to attenuate the average dislocation velocity as the overall line waits for these events to complete and the dislocation to unpin itself.

\subsubsection{The development of debris}

\begin{figure}
    \caption{Snapshots of a moving $\frac{1}{2}[111]$ screw dislocation
  (black line) projected onto $(\bar 110)$ (upper line) and $(11\bar2)$ (lower line) planes. The red lines indicate trailing debris. Blue dots represent the positions of
  hydrogen atoms. $\tau=100$~MPa, $T=300$K (a) $\CH=0$, (b) $\CH=10$~appm,  (c) $\CH=30$~appm,  (d) $\CH=50$~appm }
    \centering
    \includegraphics[scale=1,viewport=0 450 248 794,clip]{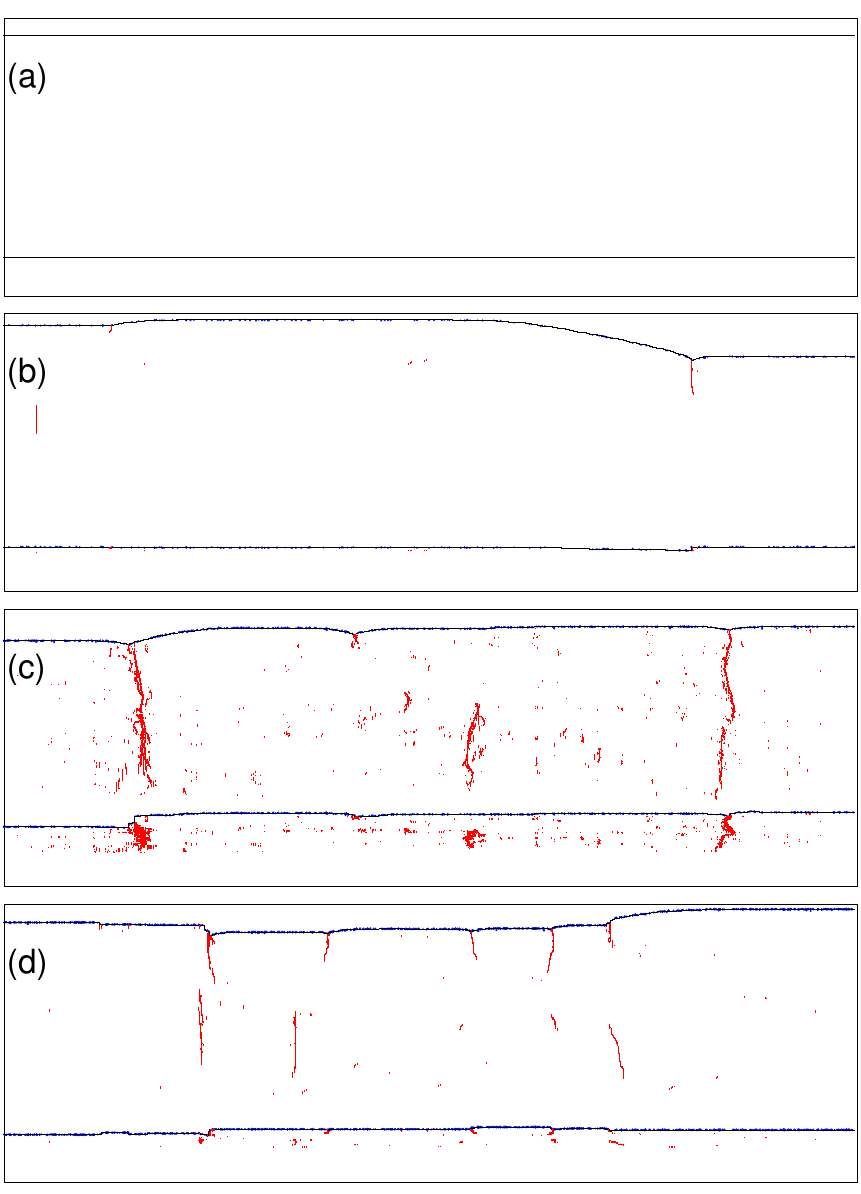}
    \label{fig:Debris}
\end{figure}

These observations are illustrated in Fig.~\ref{fig:Debris} which show simulations at a resolved shear stress, $\tau=100$~MPa, and $T=300$K. In each panel the upper black line shows a snapshot of a moving $\frac{1}{2}[111]$ screw dislocation projected onto the primary $(\bar 110)$ glide plane, while the lower black line shows the \textit{same} dislocation at the same time projected onto the perpendicular $(11\bar2)$ plane. This second projection serves to indicate the extent to which the dislocation deviates from its primary glide plane into the two cross slip planes in the $[111]$ zone. 
\begin{description}[topsep=0pt, partopsep=0pt, itemsep=0pt, listparindent=12pt, itemindent=12pt, parsep=0pt, leftmargin=6pt, partopsep=12pt, labelwidth=12pt]
\item [(a)] At $\CH=0$, panel~(a) illustrates the point made earlier that kink velocity is high and the dislocation moves as a straight line (although at $T=400$K kink pair generation is sufficiently frequent that kink collisions do occur and some debris is observed~\cite{KPP2017}). 
\item [(b)] At $\CH=10$~appm, Fig~\ref{fig:Debris}(b), $\Ekp$ is large (Fig.~\ref{fig:EkpVelStress}) and nucleation on cross slip planes is rare so that kink collisions on different slip planes is less likely---some debris is seen and the dislocation is not straight in its primary slip plane, however deviation onto a cross slip plane is limited. 
\item [(c)] At $\CH=30$~appm $\Ekp$ is small (Fig.~\ref{fig:EkpVelStress}) and nucleation on cross slip planes is commonplace: there is much debris observed and significant cross slip of the dislocation onto secondary glide planes. 
\item [(d)] As $\CH$ is further raised to 50~appm, Fig~\ref{fig:Debris}(d), $\Ekp$ is raised again (Fig.~\ref{fig:EkpVelStress}), the equilibrium core structure is the hard core and kink pair generation on the cross slip planes is again less common---less debris accumulates than at $\CH=30$~appm. 
\end{description}

\subsubsection{The ``30~appm anomaly'' and comparison with experiment}

\begin{figure}
    \caption{Calculated dislocation velocity, $\vdis$, averaged over resolved shear stresses in the interval 50--200~MPa}
    \centering
    \includegraphics[scale=0.5,viewport=65 174 555 638,clip]{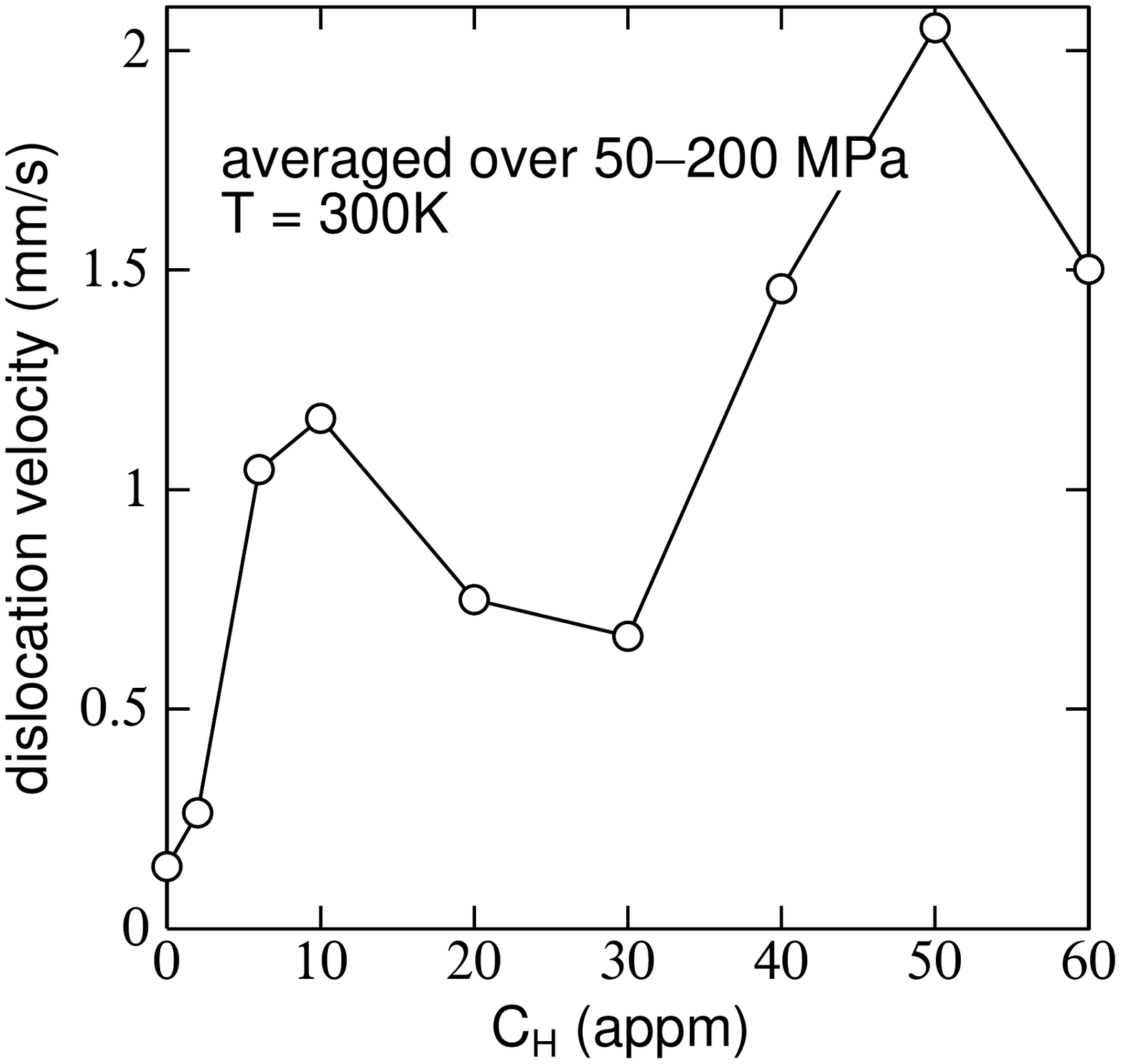}
    \label{fig:AverageVelocity}
\end{figure}

\begin{figure}
    \caption{Measured components of the activation volume for plastic shear in hydrogen loaded pure \aFe. After Wang \textit{et al.}~\cite{Wang2013}}
    \centering
    \includegraphics[scale=0.5,viewport=61 178 558 640,clip]{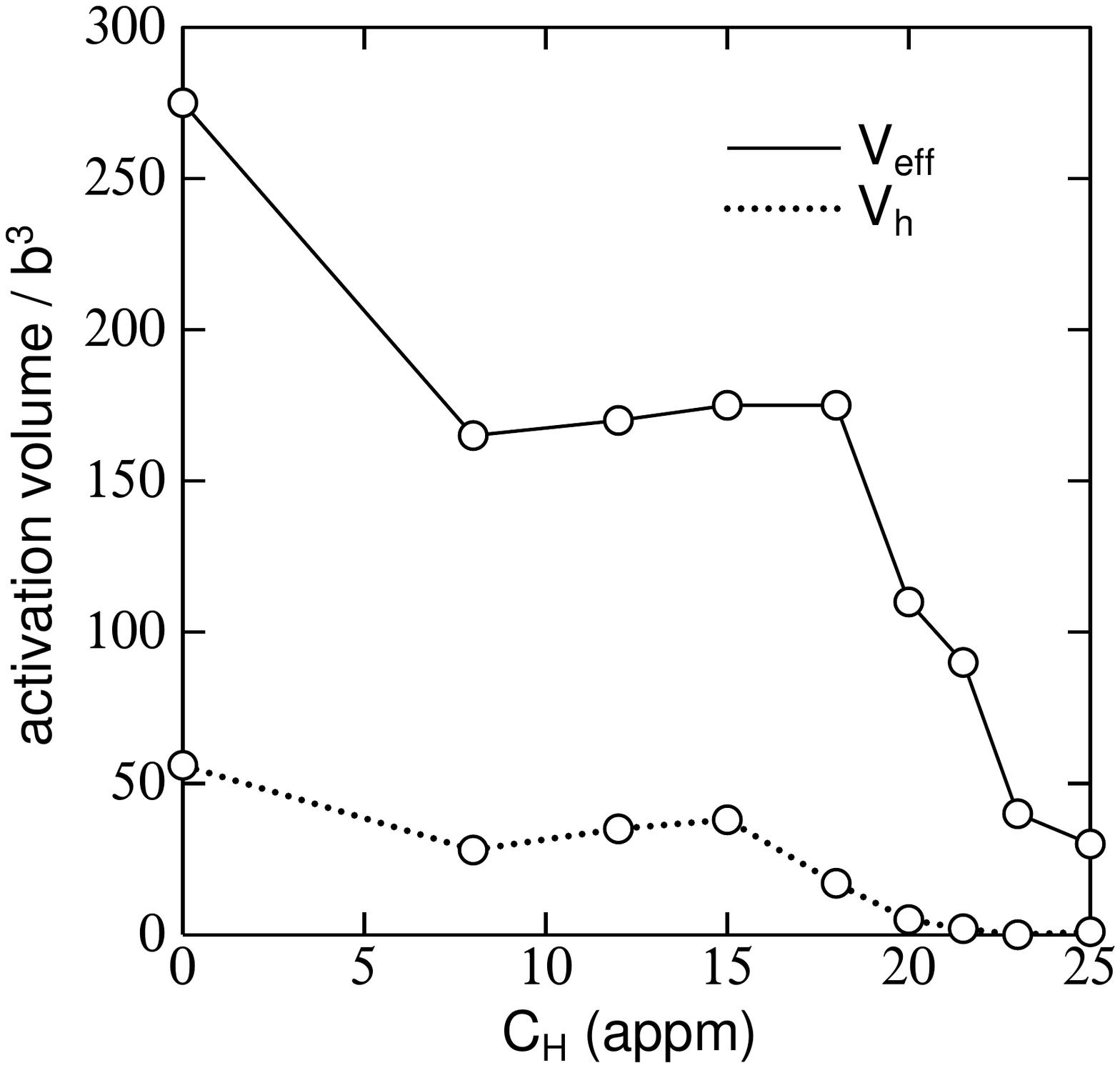}
    \label{fig:ActivationExperiment}
\end{figure}

It is very clear from all the results presented above that there is a strong non monotonic dependence of $\vdis$ on $\CH$ with an ``anomaly'' occurring around $\CH=30$~appm. The reason for this is the reduction in kink pair formation enthalpy and the associated core transformation from EC to HC. This effect is revealed most simply in Fig.~\ref{fig:AverageVelocity} which is a plot of $\vdis$ averaged over resolved shear stresses in the interval 50--200~MPa. This evident dip in $\vdis$ is mirrored in measurements of the components of the activation volume for tensile deformation of hydrogen charged \aFe. The method used is stress relaxation~\cite{Spatig1993}. The applied shear stress is divided into a thermally activated contribution, $\tau_{\rm eff}$, and a term, $\tau_\mu$, that depends on temperature only through the $T$-dependence of the shear modulus~\cite{Seeger1957},
\begin{equation*}
    \tau_{\rm app} = \tau_\mu + \tau_{\rm eff}
\end{equation*}
The strain rate as a function of temperature is given in terms of an activation free energy, $G$, of the strain rate, $\dot\gamma$, defined through,
\begin{equation*}
    \dot\gamma = \dot\gamma_0\,e^{-G/kT}
\end{equation*}
where $\dot\gamma_0$ is given by the Orowan equation~\cite{Spatig1993} and depends on the average dislocation velocity. The ``effective'' activation volume is
\begin{equation*}
    V_{\rm eff} = -\frac{{\rm d}G}{{\rm d}\tau_{\rm eff}}
\end{equation*}
What is measured is the consequence of the total applied shear stress, namely an ``apparent'' activation volume,
\begin{equation}
    V_{\rm app} = V_{\rm eff}\left(1+S'\frac{{\rm d}\tau_\mu}{{\rm d}\gamma}\right) = V_{\rm eff} + V_{\rm h}
    \label{eq_actvol}
\end{equation}
where $S'$ is the compliance of the specimen plus loading train in the tensometer. Stress relaxation tests allow the two terms, the effective and the ``hardening'' activation volumes to be identified separately. Fig.~\ref{fig:ActivationExperiment} shows such measurements, taken from Wang \textit{et al.}~\cite{Wang2013}. Hydrogen concentration is difficult to measure and coupled with uncertainties in our parameterisation of the SCkMC, there is a factor of two discrepancy between theory and experiment in the $\CH$ value where the anomaly occurs. 

It is notable that the anomaly is associated with a \textit{decrease} in the Peierls barrier and a \textit{decrease} in both the average dislocation velocity \textit{and} the effective activation volume (or, at least, a plateau in $V_{\rm eff}$). We resolve this apparent contradiction as follows. Since at this \h\ concentration the Peierls barrier and kink pair formation enthalpy is close to zero, one would \textit{expect} that $\vdis$ would be fast. However because of the peculiar three fold core structure of screw dislocations in \aFe, the fact that $\Ekp$ is nearly vanishing implies that kink pair generation is very rapid on both the primary glide plane \textit{and} the two cross slip planes. This vastly increases the likelihood of kink collisions on dissimilar slip planes leading to frequent creation of jogs and superjogs, the generation of debris and the subsequent reduction in the \textit{average} dislocation velocity. Because the thermal activation barrier is small this also reflects on the activation volume which consequently also reached a minium or plateau as seen in Fig.~\ref{fig:ActivationExperiment}.

\section{Experimental}

\subsection{Materials and Methods}

99.99\% pure iron was purchased from Goodfellow, Cambridge. Samples were $100\times 100$~mm by 2mm thick.  Specimens for stress relaxation tests were manufactured from these a tensile test piece length of 56 mm, a width in the guage area of 3 mm and a guage length of 12.5 mm. 
Hydrogen charging was undertaken using standard electrochemical techniques. The stress relaxation test specimens were charged using 1~g/L in an aqueous solution of 3~wt\% NaCl and 0.3 wt\% NH$_4$SCN with a current density of 10~mA~cm$^{-2}$ for 48~hours at room temperature. Using thermal desorption spectroscopy we determine the \h\ concentration to be 30~appm. Following charging, samples were immediately subject to repeat stress relaxation tests. Tests were undertaken using a Zwick (BTC T1-FR020 TN A50) universal testing machine. Testing was conducted under displacement control, with a strain rate of $10^{-5}$~s$^{-1}$. The specimen was initially subjected to a strain that just gave yielding of the sample. At this point, the strain was held constant for 30~s allowing stress relaxation. Subsequently the specimen was deformed to give the same as in the previous stress relaxation, and then the strain was held constant again for 30~s. The same cycles were repeated until no relaxation was recorded in the relaxation stage. Stress relaxation data was analysed to determine the values of $V_{\rm eff}$ and $V_{\rm h}$~(\ref{eq_actvol}), for the charged and uncharged specimens. We found $V_{\rm eff}=133b^3$ and $V_{\rm h}=11b^3$ and $V_{\rm eff}=127b^3$ and $V_{\rm h}=13b^3$ in uncharged and charged pure \aFe\ respectively.
In order to observe dislocation microstructures TEM thin foils were extracted from the stress relaxation tests using the FIB lift-out technique with the FEI Helios Nanolab 650 SEM/FIB instrument. During the FIB lift-out from the fracture surface, platinum was slowly deposited on the location of interest to preserve the corresponding fracture surface at the location and the microstructure below it. TEM observation of the FIB samples were then conducted in the JEOL F200 TEM operated at an accelerating voltage of 200~kV. In all cases, the imaging was carried out in STEM mode so as to show the highest dislocation density. 

\subsection{Experimental Results}
\begin{figure}
    \caption{Bright field STEM image showing the hydrogen charged, strain free state.}
    \centering
    \includegraphics[scale=1.5]{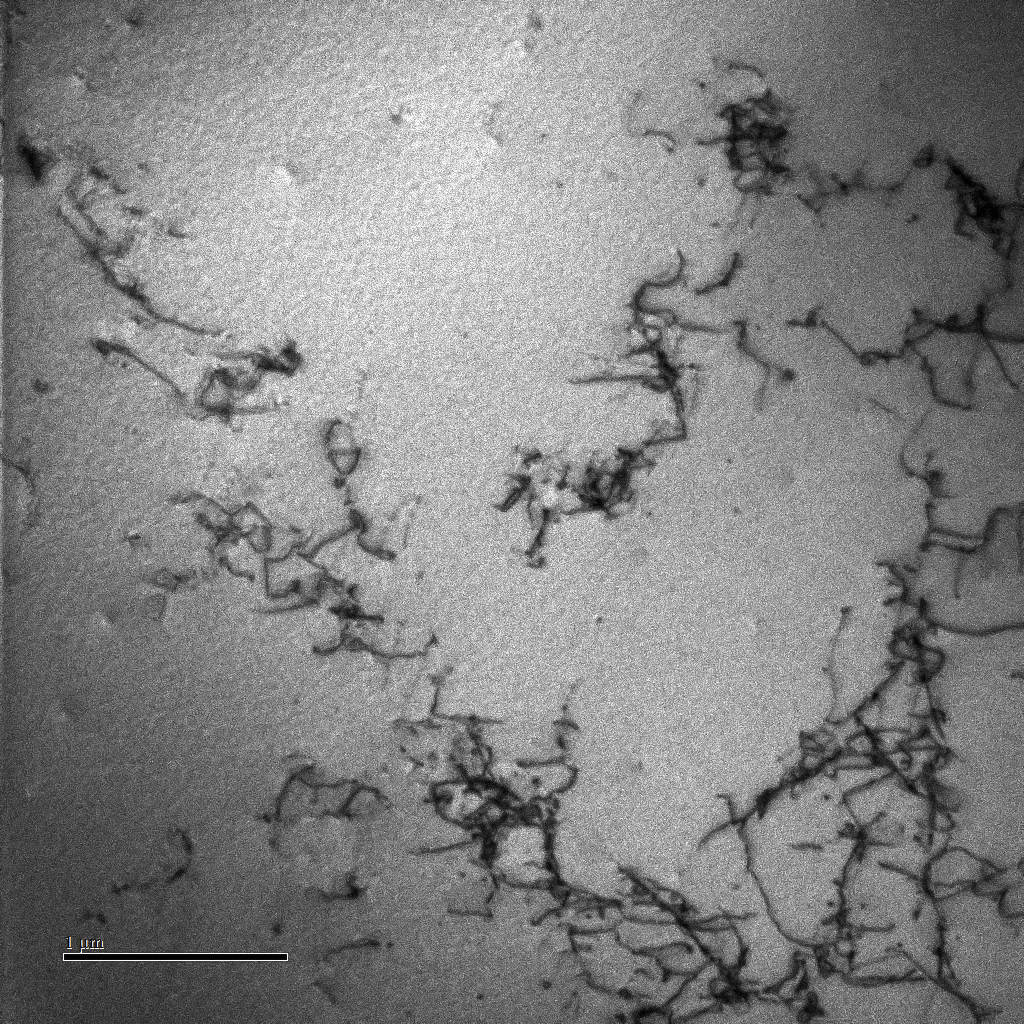}
    \label{fig:strainfree}
\end{figure}

\begin{figure}
    \caption{Dark field and bright field STEM images showing the same region of a hydrogen free specimen after tensile stress relaxation testing. At this magnification the developed cellular structure is evident as a cell wall of tangled dislocations separating dislocation free regions to left and right.}
    \centering
    \includegraphics[scale=0.75]{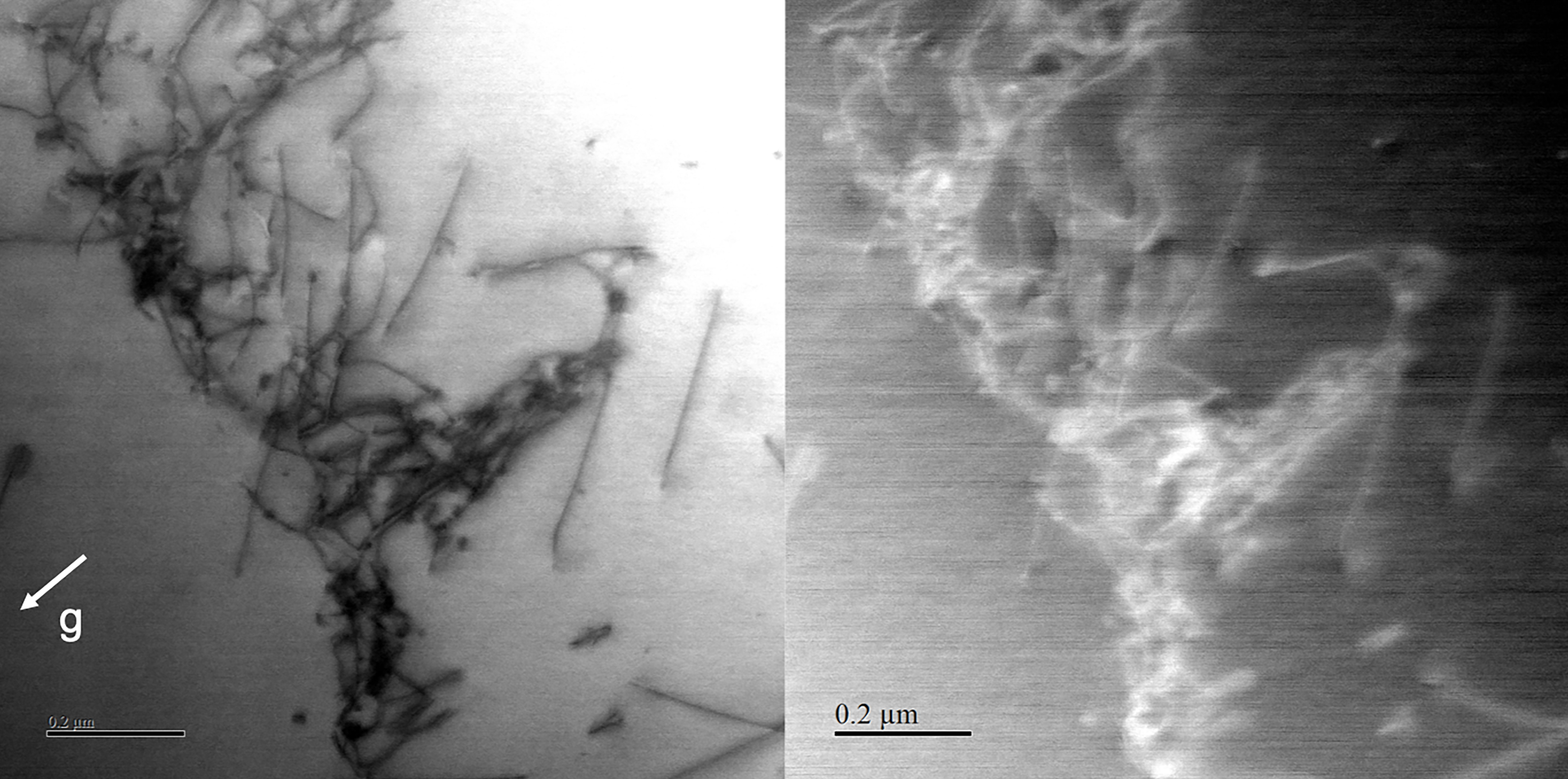}
    \label{fig:uncharged}
\end{figure}

\begin{figure}
    \caption{Dark field and bright field and dark field STEM images of the same region of a specimen \h\ charged to 30~appm after tensile stress relaxation. At the top centre can be seen a long trailing dipole. If this is in edge orientation then it may find it difficult to unzip. Another instance of self pinning into a V-shape is evidences at the top left. Multiple examples of loop debris is clearly observed.}
    \centering
    \includegraphics[scale=0.925]{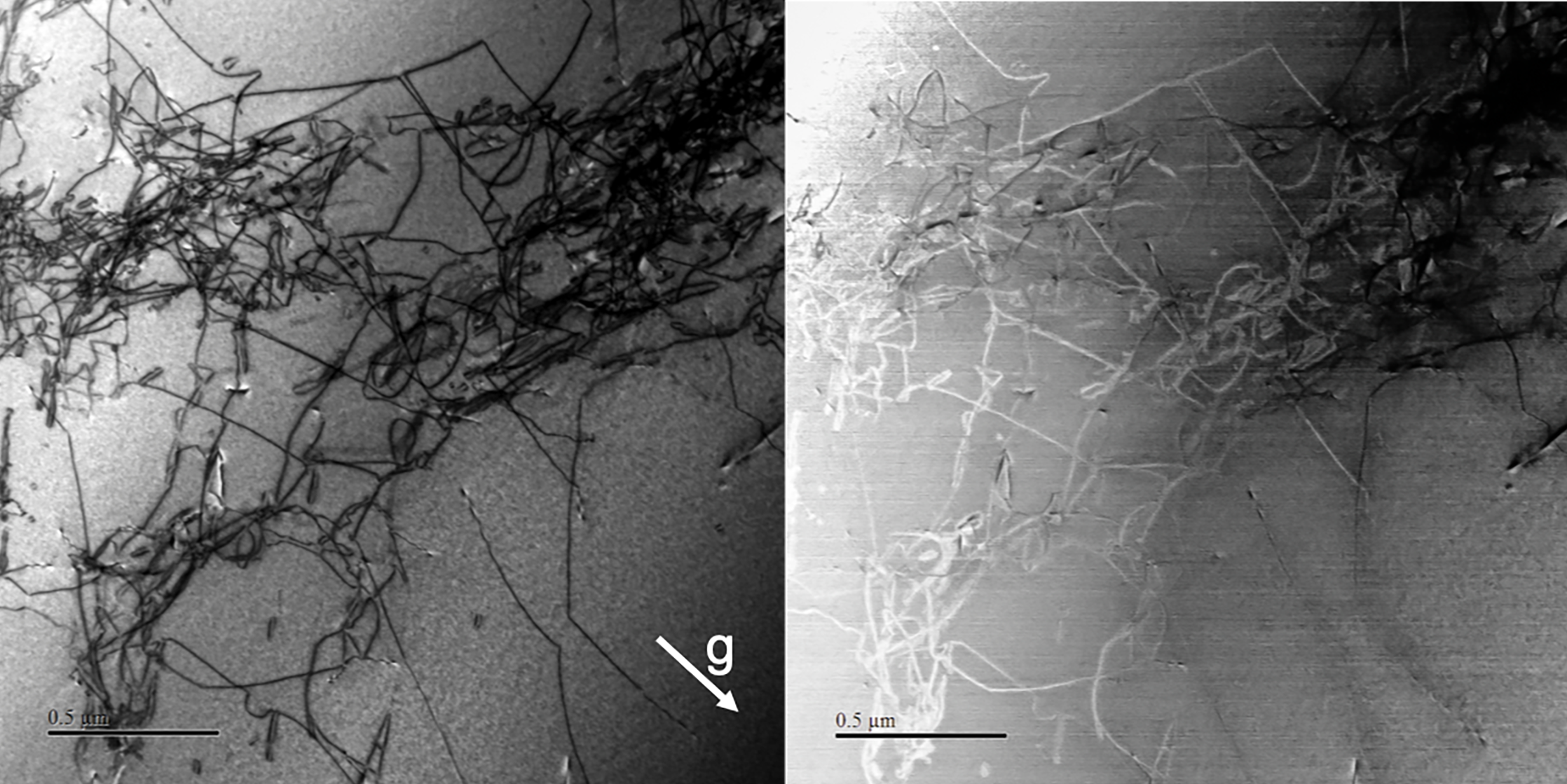}
    \label{fig:jogs}
\end{figure}

\begin{figure}
    \caption{Dark field and bright field and dark field STEM images of the same region of a specimen \h\ charged to 30~appm after tensile stress relaxation. The dislocations are rather clearly jogged and there are many loops.}
    \centering
    \includegraphics[scale=1.05]{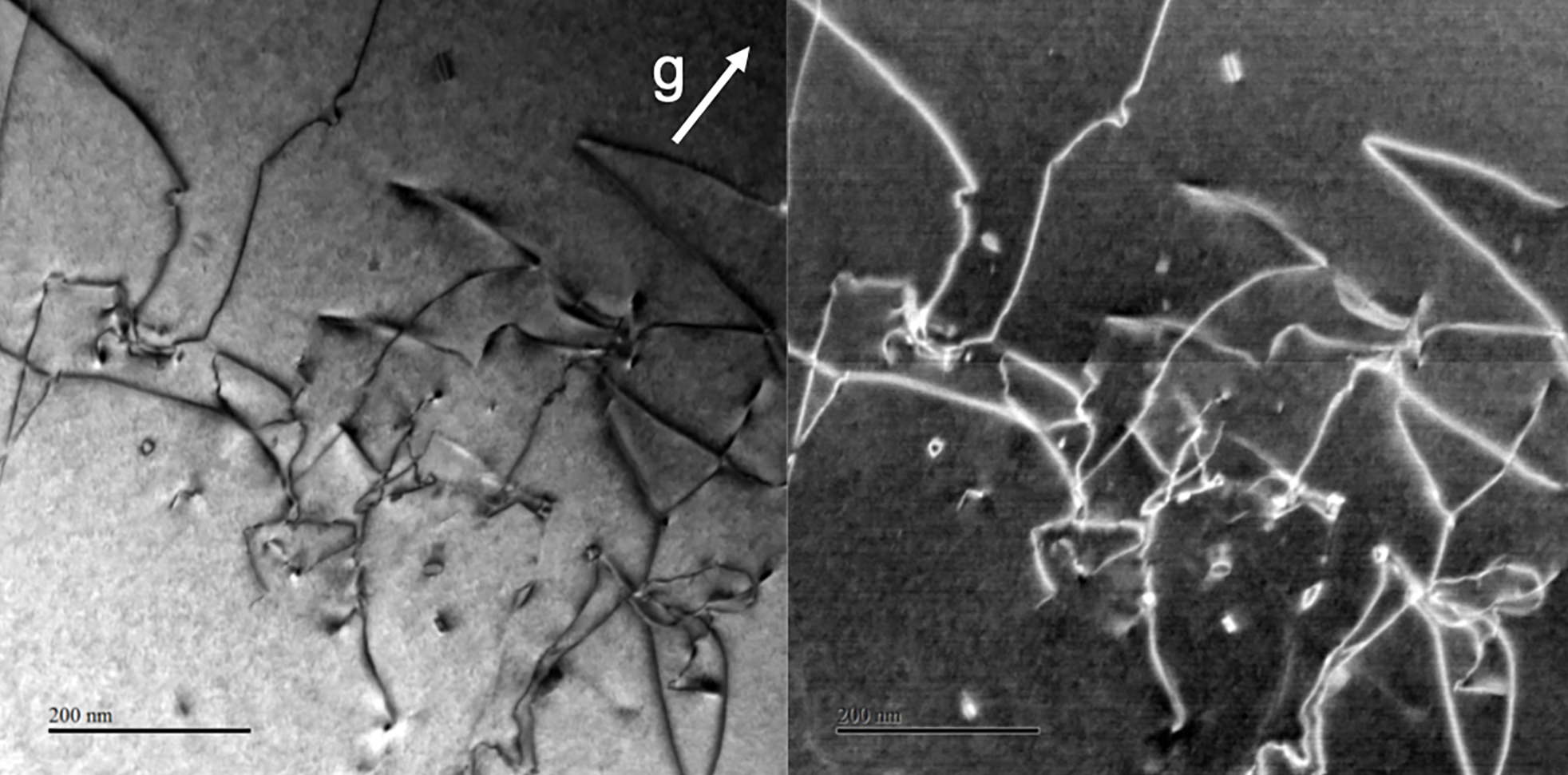}
    \label{fig:loops}
\end{figure}

Scanning transmission electron microscopy (STEM) was used to characterise the dislocation structures before and after charging in the unstrained state, and then after stress relaxation tests, again for charged and uncharged specimens. All images were recorded under the same two-beam, $g = [110]$, conditions in order to ensure that the dislocation structures could be directly compared. The dislocation structures in the hydrogen free, strain free samples are as expected, with a low dislocation density, comprising largely homogeneous dislocation distributions. On charging, but without strain, the dislocation density measurably increases, although the total dislocation density remains low. The dislocations tend to be in tangles forming rudimentary cell walls, Fig.~\ref{fig:strainfree}, with the results entirely consistent with those of Wang \textit{et al.}~\cite{Wang2013}.
The change in dislocation structure following stress relaxation testing of the hydrogen free pure iron samples is shown in Fig.~\ref{fig:uncharged}. Dislocations are arranged into a rudimentary cell structure, with individual dislocation lines easily imaged in places. The dislocation structures for the hydrogen charged stress relaxation samples are markedly different. Dense dislocation tangles are present, Fig.~\ref{fig:jogs}. There are numerous examples of jogs and dislocation debris such as prismatic loops. This result is consistent with Fig.~\ref{fig:loops}. As noted before, the dislocation jogs act as self-pinning points, which result in edge dipoles being dragged out, leading to a train of prismatic loops. This effect is purely a result of the dissolved hydrogen in the sample. 


\section{Discussion}

\subsection{Interpretation of activation volume and the ``30 appm anomaly''}

Our calculations of average dislocation velocity as a function of \h\ concentration, Fig.~\ref{fig:AverageVelocity}, show a deep minium at about 30~appm where the self pinning of the dislocation attenuates the otherwise increasing $\vdis$ with $\CH$ and almost returns $\vdis$ to that of pure \aFe. This is reflected in the measured effective activation volume from Wang \textit{et al.}~\cite{Wang2013} and plotted in Fig.\ref{fig:ActivationExperiment}. In this plot, $V_{\rm eff}$ is seen to rise at about 18~appm, but not quite to reach that of pure \aFe\ before falling to a small value associated with enhanced $\vdis$ due to \h. Our own measurements of effective activation volume show that at 30~appm of \h\ that of the charged specimen, $127b^3$, is only a little smaller than the uncharged specimen, $133b^3$. This is fully consistent with our calculations which show that $\vdis$ decreases to approach that of pure \aFe\ at the ``30~appm anomaly''. The only discrepancy with the data of Wang \textit{et al.}~\cite{Wang2013}, Fig.\ref{fig:ActivationExperiment}, is that they find the ``anomaly'' at about 18~appm. There are uncertainties in measurement of $\CH$; and the values of $\CH$ appropriate to our simulations are set by the trap energies deduced from DFT whose errors are amplified exponentially in the McLean isotherm~(\ref{eq_McLean}). We are therefore not disturbed by a factor of nearly two discrepancy between our work and that of Wang \textit{et al.}~\cite{Wang2013}. We suggest that the ``anomaly'' is real and will have great significance in both the interpretation of experiments and in the establishment of non trivial models for dislocation velocity to be used in multiscale models of \h\ embrittlement. A key finding here is that the \textit{macroscopic} measurements that indicate the anomaly can be traced \textit{microscopically} to the screw dislocation core transformation brought about by \h, Fig.~\ref{fig:LineTensionHydrogen}.

\subsection{Implication for dislocation cell formation}

\begin{figure}
    \caption{The angle describing the deviation of a long straight $\frac{1}{2}[111]$ screw dislocation from the primary glide plane, after moving
 over a distance of 50~nm, as a function of $\CH$ and resolved shear stresses.}
    \centering
    \includegraphics[scale=0.5,viewport=7 13 436 334]{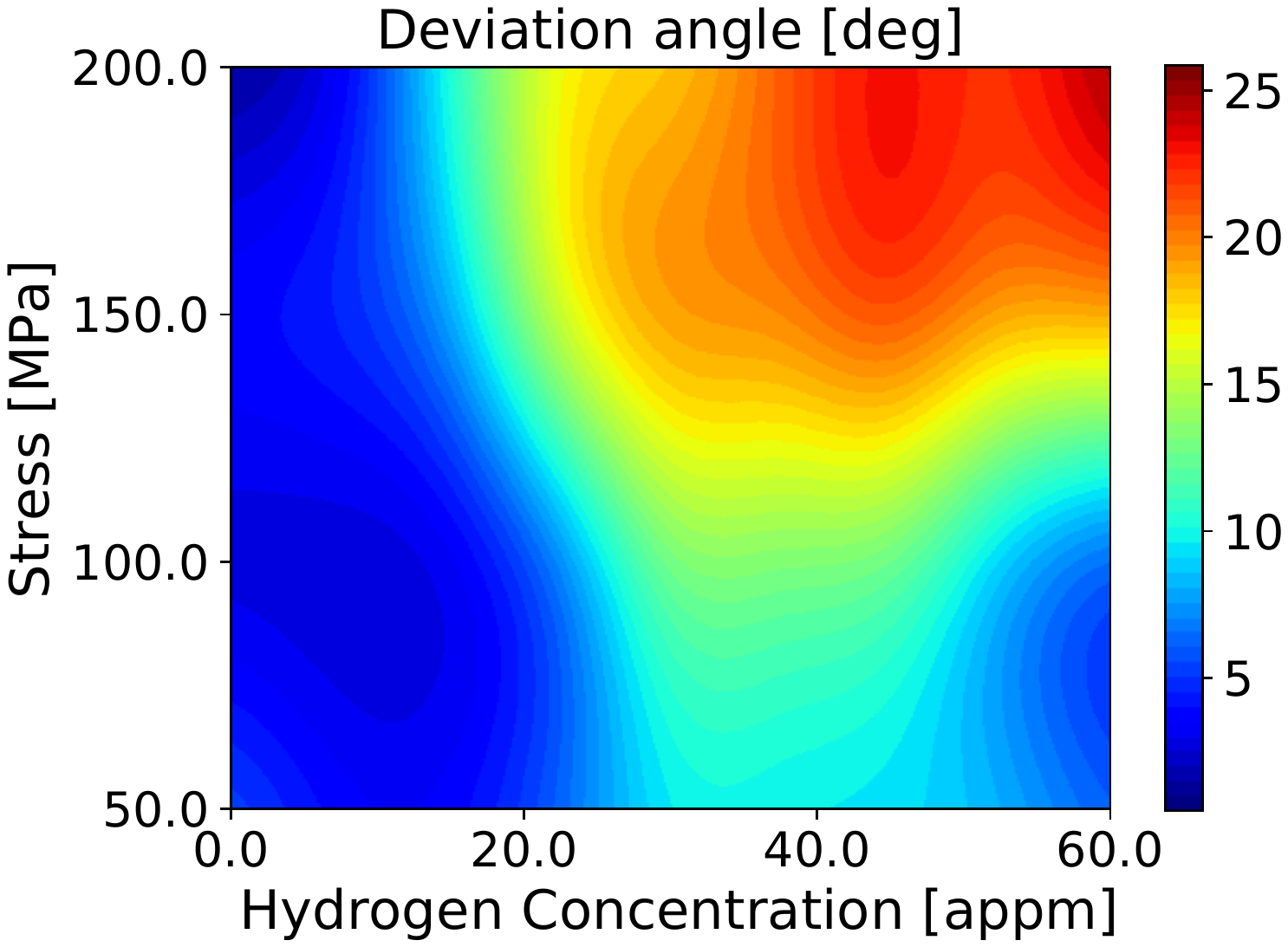}
    \label{fig:DeviationAngle}
\end{figure}

Our transmission electron microscopy observations of dislocation structures of both \h\ charged and uncharged specimens in \aFe\ show that the homogeneous dislocation forest existing in \h\ free samples transforms into cell walls that separate relatively dislocation free regions.  The cell walls can be regarded as dense dislocation tangles. The driving force for cell wall formation arises from the reduction in the total
elastic energy of the dislocations due to their clustering.  TEM images have shown that the volume of the dislocation free zones and the density of the tangled structures increase with increasing $\CH$ in the interval 0--25~appm~\cite{Wang2013}. The physics that lies behind dislocation reorganisation due to hydrogen is not yet well understood. It has been commonly accepted that a requirement for cell formation is
that dislocations have sufficient mobility out of their slip plane \cite{Hirth1959}. Therefore, whether cells form or not depends on factors which determine the ease with which dislocations cross slip or climb. The present SCkMC simulations and experiments show that the probability for formation of
kink pairs in secondary slip planes and dislocation segments which glide out of the primary slip plane increases with $\CH$. The angle describing the deviation of the dislocation
from the primary slip plane as a function of applied stress and $\CH$ is shown in Fig.~\ref{fig:DeviationAngle}.
Again, at moderate stresses where glide is dominated by kink pair formation enthalpy, we see an anomaly near $\CH=30$~appm near the EC--HC core transformation, where $\Ekp$ is small and kink pair activation is prolific on the cross slip planes.
SCkMC simulations show that the dislocation mobility out of the primary slip plane increases significantly for $\CH > 10$~appm and applied stresses higher than 100~MPa. This result of the SCkMC simulations agrees with TEM observations indicating an increase of the density of the tangled structures with increase of hydrogen concentration.

It is a matter of further work to investigate the role of \h\ and other interstitials such as carbon on the dynamics of cell formation and the development of cellular and sub-grain microstructures.

\section{Conclusions}

\begin{description}[topsep=0pt, partopsep=12pt, itemsep=0pt, listparindent=12pt, itemindent=12pt, parsep=0pt, leftmargin=6pt, partopsep=12pt, labelwidth=12pt, labelsep=0pt]
\item[1.] We demonstrate a new self consistent kinetic Monte Carlo scheme that is able to calculate average dislocation velocity of long straight $\frac{1}{2}[111]$ screw dislocations in pure and \h\ loaded \aFe. The self consistency arises because of the parametric dependences of the speed on the kink pair formation enthalpy, and the kink pair formation enthalpy on the speed.
\item[2.] As predicted in previous work~\cite{KPP2017}, we find that an effect of \h\ is to generate large quantities of debris behind a moving screw dislocation, even at room temperature. This is a consequence of kink collisions on different slip planes. 
\item[3.] The predicted debris has now been found in TEM images of \h\ loaded \aFe, following tensile stress relaxation testing.
\item[4.] We have identified what we call the ``30~appm anomaly''. This corresponds to the \h\ concentration at which there is a core transformation of the screw dislocation from easy core to hard core configuration. At the critical concentration, the Peierls barrier and kink pair formation enthalpy approach close to zero, before increasing as $\CH$ increases beyond 30~appm and the barrier for glide subsequently appears at the EC state. Signatures of the anomaly are, (\textit{i\/}) very frequent kink pair production and creation of self pinning jogs, (\textit{ii\/}) a \textit{minimum} in $\vdis$ due to the profilific creation of debris (see Figs.~\ref{fig:Debris},\ref{fig:AverageVelocity}) (\textit{iii\/}) a plateau, or minimum in the effective activation volume for slip (Fig.~\ref{fig:ActivationExperiment}).

\end{description}

\section*{Acknowledgements}
We acknowledge the support of EPSRC under the Programme Grant HEmS, EP/L014742. 
IHK acknowledges support from Bulgaria National Science Fund (BNSF) under Programme grant KP-06-H27/19. 


\bibliography{help}

\end{document}